\documentclass[conference]{IEEEtran}
\usepackage{graphicx}
\usepackage{multirow}
\PassOptionsToPackage{hyphens}{url}\usepackage{hyperref}
\usepackage{url}
\usepackage{subcaption}
\usepackage{calc}
\usepackage{amsthm}
\usepackage{amsmath}
\usepackage{algorithm}
\usepackage[noend]{algpseudocode}
\usepackage{algorithmicx}
\usepackage{xcolor}
\usepackage{multirow}
\usepackage{fancyhdr}
\begin{document}

\title{A Model Drift Detection and Adaptation Framework for 5G Core Networks}
%
%
%

\author{Dimitrios~Michael~Manias, Ali~Chouman, and Abdallah~Shami\\
ECE Department, Western University, London ON, Canada\\
\{dmanias3, achouman, Abdallah.Shami\}@uwo.ca}


\maketitle

\begin{abstract}
The advent of Fifth Generation (5G) and beyond 5G networks (5G+) has revolutionized the way network operators consider the management and orchestration of their networks. With an increased focus on intelligence and automation through core network functions such as the NWDAF, service providers are tasked with integrating machine learning models and artificial intelligence systems into their existing network operation practices. Due to the dynamic nature of next-generation networks and their supported use cases and applications, model drift is a serious concern, which can deteriorate the performance of intelligent models deployed throughout the network. The work presented in this paper introduces a model drift detection and adaptation module for 5G core networks. Using a functional prototype of a 5G core network, a drift in user behaviour is emulated, and the proposed framework is deployed and tested. The results of this work demonstrate the ability of the drift detection module to accurately characterize a drifted concept as well as the ability of the drift adaptation module to begin the necessary remediation efforts to restore system performance. 
\end{abstract}

\begin{IEEEkeywords}
Model Drift, Concept Drift, 5G Core Networks, Next-Generation Networks, NWDAF
\end{IEEEkeywords}

\section{Introduction}
The introduction of 5G networks has caused a significant paradigm shift in the networking landscape. With new applications and use cases being constantly developed (\textit{i.e.,} ultra-reliable low latency communications, massive machine-type communications, and enhanced mobile broadband \cite{ ETSI_NWDAF}), network operators are required to accommodate such applications and ensure adequate levels of QoS are attained. With the increase in global connectivity demands and the introduction of IoT and M2M communications, 5G+ networks and systems will have massive amounts of network-generated data. Considering the speed at which this data is being produced and consumed, along with the variety of data collected, 5G+ systems are classified as big data applications \cite{zheng2016big}. Given the increasing complexity of these systems, intelligence has been highlighted for its ability to provide enhanced network management and orchestration capabilities \cite{manias2020need}.

The latest 5G specification identifies the Network Data Analytics Function (NWDAF) as a key enabler for intelligence in 5G networks \cite{3gppR16}. Through this core network function, increasing system intelligence and automation levels will be attained. While the NWDAF offers a fundamental first step towards total network intelligence and automation, 5G+ networks are expected to have a much more profound integration of intelligent systems and agents beyond a single core network function. In order to reach this stage, one of the major challenges of the widespread implementation and adoption of intelligent agents, model drift, must be solved.

Model drift is a phenomenon which plagues machine learning implementations and can manifest itself through concept drift and data drift. Concept drift defines a fundamental change in the underlying relationship or process, whereas data drift, as its name suggests, describes a change in the observed data distribution without a change in the underlying relationship \cite{manias2021}. An example of concept drift in 5G and beyond networks is changing user habits, such as introducing a new application or platform. An example of data drift can be an incomplete or partial representation of a population, such as an intrusion detection system that encounters a new threat -- the fundamental relationship between a threat and non-threat has not changed; however, the set of threats has been expanded and was previously misrepresented. The work presented in this paper will consider a concept drift whereby various classes of User Equipment (UEs) are exposed to new applications being periodically introduced. The contributions of the work presented in this paper can be summarized as follows:
 \begin{itemize}
     \item The emulation of a concept drift event in a functional 5G core prototype and the collection of the associated network traffic data.
     \item The development of a drift detection framework for 5G core networks, which is able to accurately identify drifted concepts in user behaviour.
     \item The development of an online drift adaptation framework capable of remediating the impact of a drift and progressing towards restoring the system to pre-drift performance.
 \end{itemize}
 The remainder of this paper can be summarized as follows. Section II discusses related work and presents the state-of-the-art in the field. Section III presents the methodology and implementation followed in this work, including the 5G core prototype, the experimental setup, the machine learning model implemented, as well as the drift detection and adaptation frameworks. Section IV presents and analyzes the results. Finally, Section V concludes the paper and presents opportunities for future work in the field.

\section{Related Work}
The following section will outline related work in the field of concept drift detection. Traditionally, concept drift detection has been accomplished by methods which monitor the real-time performance of a model \cite{gama2004learning} or ones which monitor the distribution of the data being provided to the model \cite{bifet2007learning}. Garcia \textit{et al.} \cite{baena2006early} extend the idea of error rate monitoring and instead consider the distance between classification errors as an indication of a drift. The work of Manias \textit{et al.} \cite{manias2021} leverages federated node model updates to perform concept drift detection in federated networked systems. Through the dimensionality reduction, clustering, and modelling of normal federated node update behaviour, the proposed framework is able to identify drifted concepts based on a distance-based threshold calculation. \par

Yang and Shami \cite{ yang2021lightweight} overview various techniques for concept drift adaptation, including adaptive algorithms, which fully retrain a model when a new concept is detected, incremental learning algorithms, which periodically resume model training in the presence of a new concept, and ensemble methods which use a variety of models built on past and present concepts to make a prediction. They propose a lightweight concept drift adaptation framework for IoT data streams in their work. Their work uses an optimized LightGBM model that is periodically retrained when a new concept is detected. Yang \textit{et al.} \cite{ yang2021} also propose an ensemble framework for concept drift adaptation for IoT streams. Their work uses a set of base learners with a dynamic weighting scheme to update each model’s contribution to the final outcome. \par

The work presented in the paper uses online performance metric monitoring coupled with the construction of a model of normal behaviour to detect a concept drift. Furthermore, two modes of operation are discussed, including persistent and non-persistent memory, which enable incremental learning to occur. When dealing with concept drift in networks, it is critical to consider the tradeoff between performance and resource utilization (\textit{i.e.,} communication, computational, \textit{etc.}); as such, this work presents a parameterized concept drift detection and adaptation framework which can be tuned to adhere to such requirements. Additionally, the work presented in this paper further differs from the state-of-the-art as it directly addresses the 5G core network, which has not been considered in previous work. One of the main challenges in machine learning and networking applications is the unavailability of high-quality public datasets due to privacy and proprietary reasons. As such, the related work in the field has a major limitation of not being evaluated on real-world network data. This is a significant shortcoming as the intricacies and complexities of network data are not present in synthetic and toy datasets. \footnote{The network data used in this work is available at \url{https://github.com/Western-OC2-Lab/5G-Core-Networks-Datasets}} 

\section{Methodology and Implementation}
\subsection{5G Prototype}
The 5G prototype used for this work is discussed in brief; however, a full description of this system is available through the previous work of Chouman \textit{et al.} \cite{chouman2022}. The 5G core prototype leverages the Open5GS \cite{Open5GS} and UERANSIM \cite{aligungr} open-source software to provide 5G core network functionality and 5G UE and RAN functionality, respectively. Through Open5GS core network functions, including the Network Repository Function (NRF), Access and Mobility Function (AMF), Session Management Function (SMF), Authentication Server Function (AUSF), Unified Data Management (UDM), Unified Data Repository (UDR), Policy Control Function (PCF), Network Slice Selection Function (NSSF), Binding Support Function (BSF), as well as the User Plane Function (UPF) are implemented. Furthermore, the NWDAF, which as of December 2021, provides multiple services, including Analytics Subscription, Analytics Info, Data Management, ML Model Provision, and ML Model Info \cite{3gppR16}, has been developed, deployed and integrated with the prototype. 
\subsection{Experiment Setup}
\begin{figure}[!htbp]
\centerline{\includegraphics[width=0.95\columnwidth]{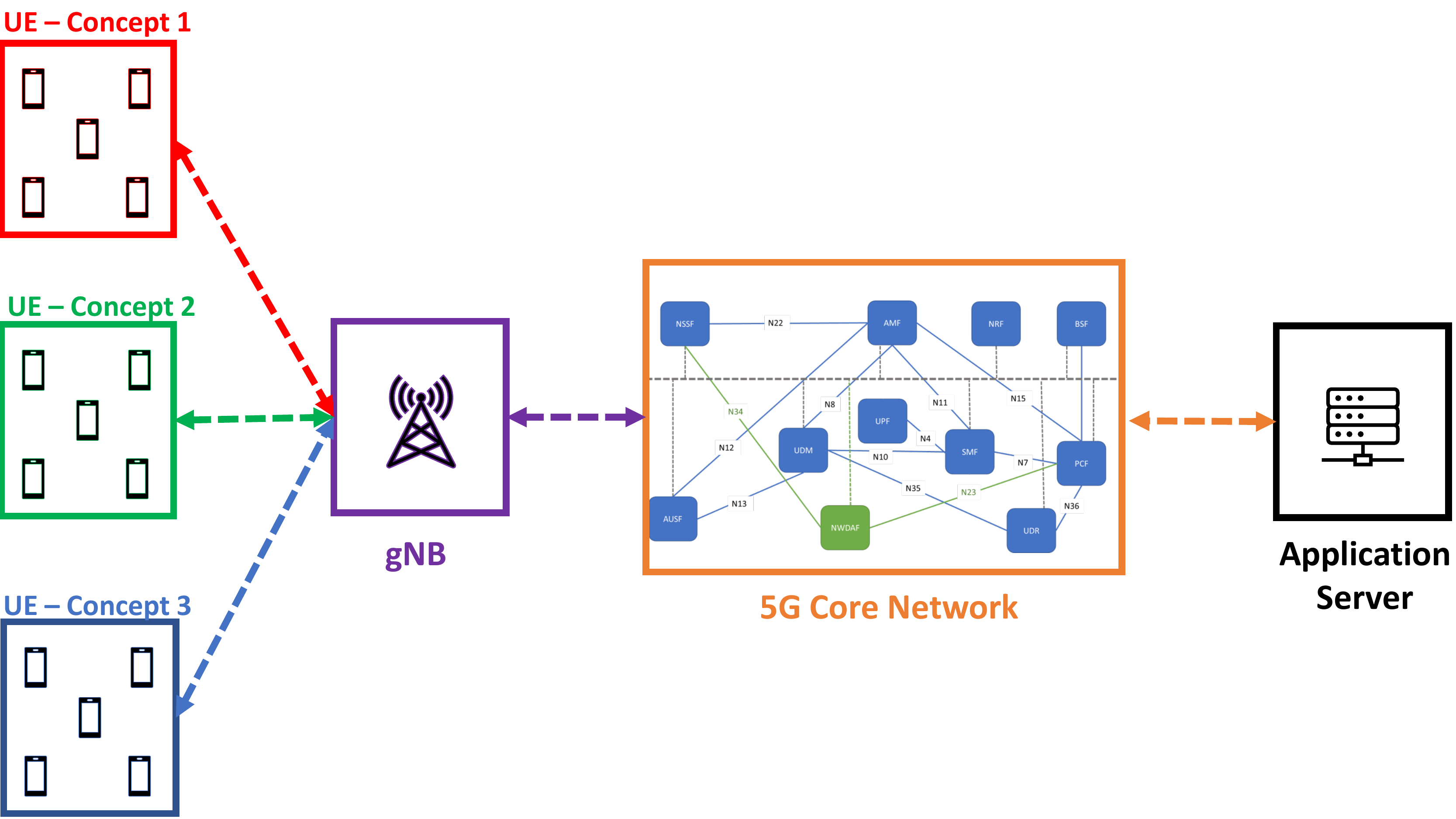}}
\caption{Experiment System Model}
\label{app}
\end{figure}

An application server is created with its associated REST APIs for the experimental setup. The server currently supports one application (image downloading); however, it will be expanded to include additional applications in the future. Hosted on the server are images of various sizes, including 16x16, 32x32, 48x48, 128x128, 256x256, and 512x512. UEs periodically request to download an image through the associated API call. Three UE classes are created corresponding to various concepts; the first class is limited to selecting images of size 16x16 and 32x32, the second class is limited to selecting images of size 48x48 and 128x128, and the final class is limited to selecting images of size 256x256 and 512x512. Additionally, the frequency of the requests differs between the classes, where class one has the greatest frequency, and class three has the lowest frequency. Initially, the system is run with one concept present; however, a new concept is periodically introduced. It should be noted that an emerging concept does not replace the former concept but rather co-exists with it. Figure \ref{app} illustrates the three classes of UEs and their communication with the Application Server through the gNB and 5G Core Network.

\subsection{Model}
The objective of the proposed model is to predict the length of the next packet sent from a UE to the application server, which can be used as part of a traffic estimation or load balancing use case. A sequence of the ten previous packet lengths is supplied to the model as the input feature set, and a single value corresponding to the next packet length prediction is returned as the output. The model is composed of the input layer receiving the input sequence followed by four hidden layers leading into a single-node output layer. The first hidden layer is an LSTM layer with 100 LSTM cells. An LSTM layer was included as the first layer in the model due to its ability to extract temporal relationships from data sequences. The remaining hidden layers are fully connected, leveraging a declining neuron scheme where hidden layer two has 75 neurons, hidden layer three has 50 neurons, and hidden layer four has 25 neurons. All hidden layers use the rectified linear unit activation function. \par
The model is compiled using the mean squared error as a loss function since we are dealing with a regression task and the Adam optimizer for its computational efficiency and low memory requirement \cite{kingma2014adam}. The model is trained for 200 epochs. The training set of the model consists of the time-shifted sequence of packet lengths obtained from a single UE-to-server interaction. The assumption made is that the selected UE belongs to the initial concept and that no drifted behaviour is present during the initial training phase. As discussed in the results, the model presents substantial evidence of learning and maintains its performance when evaluated on additional UE-to-server interactions belonging to the initial concept. 

\subsection{Drift Detection and Adaptation Framework}
\begin{figure*}[!htbp]
\centerline{\includegraphics[width=1.90\columnwidth]{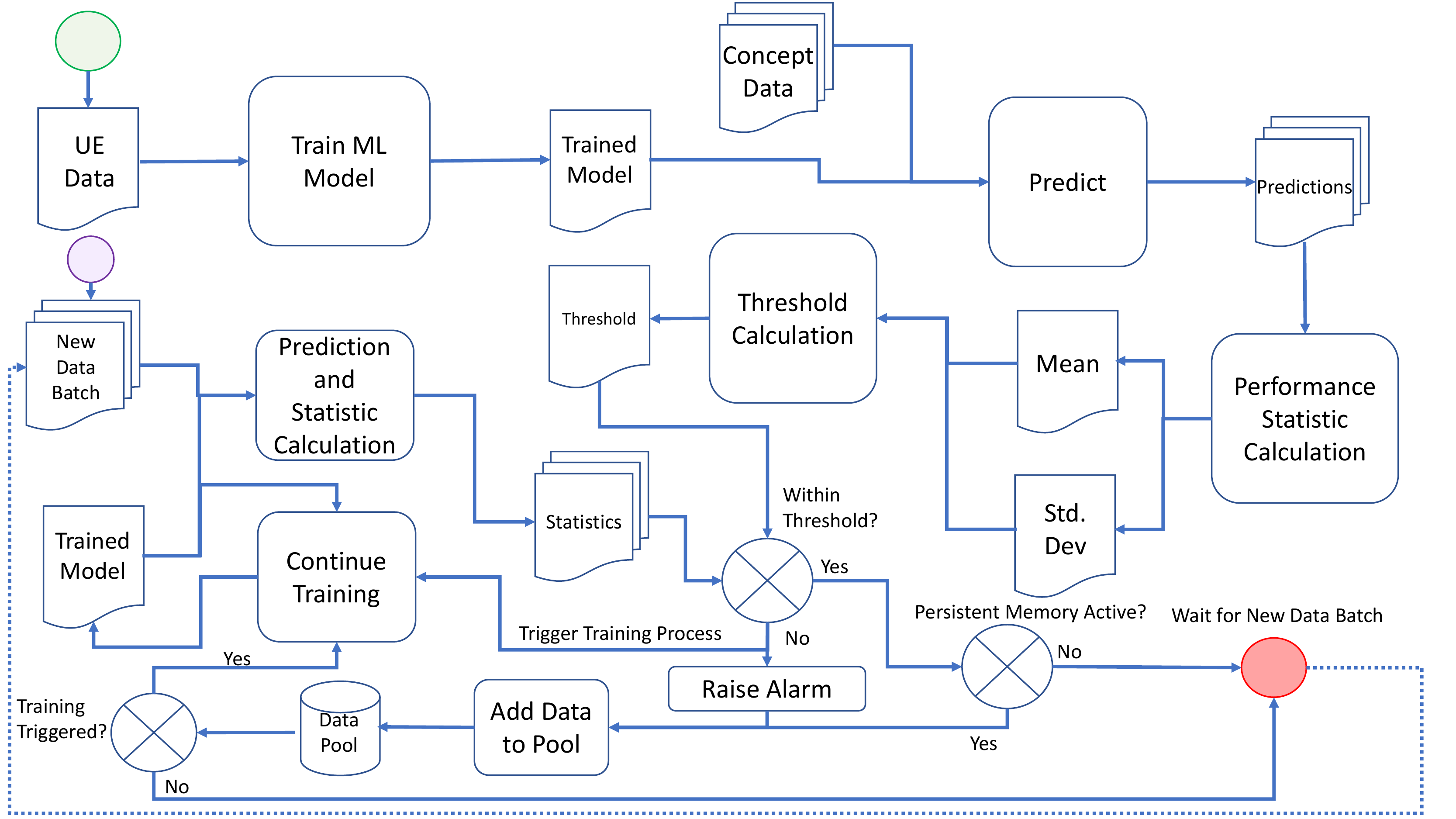}}
\caption{Drift Detection and Adaptation Framework Process Map}
\label{pmap}
\end{figure*}

The drift detection and adaptation framework is depicted in the process map of Fig. \ref{pmap}.

\subsubsection{Drift Detection}
The drift detection module is the first part of this framework and comprises the steps beginning from the green entry point with the initial UE data until the threshold calculation in Fig. \ref{pmap}. As previously mentioned, the model used in this work is trained on a single UE-to-server interaction. Once trained, the model is required to perform inference on other UE-to-server interactions, which it was not trained on. Combined with the initial training data, these UE-to-server interactions are considered concept data since the underlying assumption is that no drift events have occurred at this stage. When the predictions on the concept data are made, the performance statistics are calculated; in this case, since the model was trained using the mean squared error loss function, the squared error of each predicted value is calculated. Once calculated, the mean and standard deviation of all the individual squared error values are determined. These values are used to describe the model performance and are used to build the threshold. The threshold for this work follows Eq. \ref{thresh} where $\mu$ is the calculated mean, $\sigma$ is the calculated standard deviation, and $n$ is a positive integer used to control the conservativeness of the threshold. Since we are dealing with minimizing the mean squared error, only the right side of the interval is required; if the metric used was a maximization metric, we would use the left side of the interval. Intuitively, a smaller $n$ corresponds to a smaller threshold and, therefore, stricter performance, whereas a larger $n$ translates to a larger threshold with more lenient performance. 
\begin{equation}
\mu + n \times \sigma
\label{thresh}
\end{equation}
At this point, the drift detection module has been created and ‘trained’ and is ready to be actively deployed as part of the framework. 

\subsubsection{Drift Adaptation Module}
The drift adaptation module comprises the steps of Fig. \ref{pmap} beginning at the purple entry point. After the drift detection module has been ‘trained’, the drift adaptation module goes online. For this stage, the incoming data is grouped into batches of size $\beta$. The trained model predicts on the incoming batch of data, and then the statistic is calculated (in this case, the mean squared error of the batch). This statistic is then passed to the active drift detection module, which determines if the calculated score is within the threshold or not. A drift alarm is raised if the calculated metric is outside the threshold and the training process is triggered. The training process is parameterized by $\tau$, corresponding to the number of additional training epochs conducted. Once the training process has been completed, the trained model is used instead of the initial model. 
An additional component to the drift adaptation module must be discussed relating to the data used for the additional training iterations in the case of a drift alarm. There are two modes of operation for this module, persistent and non-persistent memory. In the event of a drift alarm being raised, as seen in the process map, the current batch of data is added to the data pool and is used to continue training. This process is common to both the persistent and non-persistent memory modes. In the non-persistent memory mode, only data batches resulting in raised drift alarms are added to the data pool. Conversely, in the persistent memory mode, $\theta$ batches after a drift alarm has been raised are added to the data pool; that is, even if a drift alarm has not been raised, data can still enter the data pool. This is depicted in the process map by the ‘Persistent Memory Active’ decision. 
If a batch of data is added to the data pool as part of the persistent memory mode without raising a drift alarm, model training has not been triggered, and therefore the process continues and waits for a new incoming batch of data. When a new batch of data is received, the process is repeated and continues running perpetually. Periodic adjustment of the threshold, especially in the presence of drift, has not been considered in this work and is an opportunity for future work in the field.

\section{Results and Analysis}
The parameterization of the drift detection and adaptation framework for the presented results is as follows: $n$ = 2, $\beta$ = 10, $\tau$ = 25, and $\theta$ = 5. The data collection phase lasted for 16 minutes, during which the experiment was conducted. The following section will outline the drift's characteristics and the framework’s performance on unseen concepts for both the persistent and non-persistent memory modes. 
\subsection{Concept Characterization}
\begin{figure*}[!htbp]
\centerline{\includegraphics[width=1.95\columnwidth]{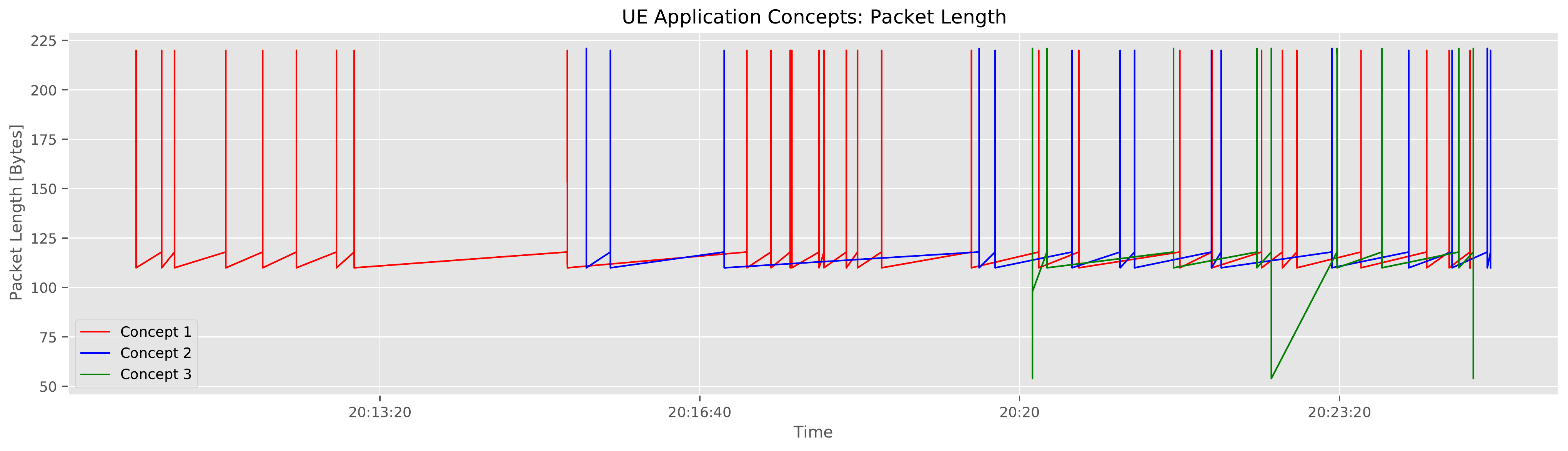}}
\caption{UE-to-Server Interaction Concepts}
\label{comparative}
\end{figure*}

As previously mentioned, there are three concepts present in this data, each relating to a specific type of user behaviour. The main method of concept characterization involves exploring the UE-to-server interactions. Figure \ref{comparative} highlights the sending of packets from the UE to the server for three UEs, each belonging to a different concept. As seen in this figure, in line with the experiment description, only concept one was initially present, then concepts two and three were introduced periodically. From this figure, it is evident that concept one is closest to concept two in terms of the pattern of packet transmission; however, it is significantly different in terms of transmission frequency. Furthermore, concept three is the most distant from the other two concepts both in terms of transmission pattern and frequency. As such, it is expected that adjusting and adapting to concept two will be easier than adapting to concept three. 

\begin{figure*}[!htbp]
\centerline{\includegraphics[width=1.95\columnwidth]{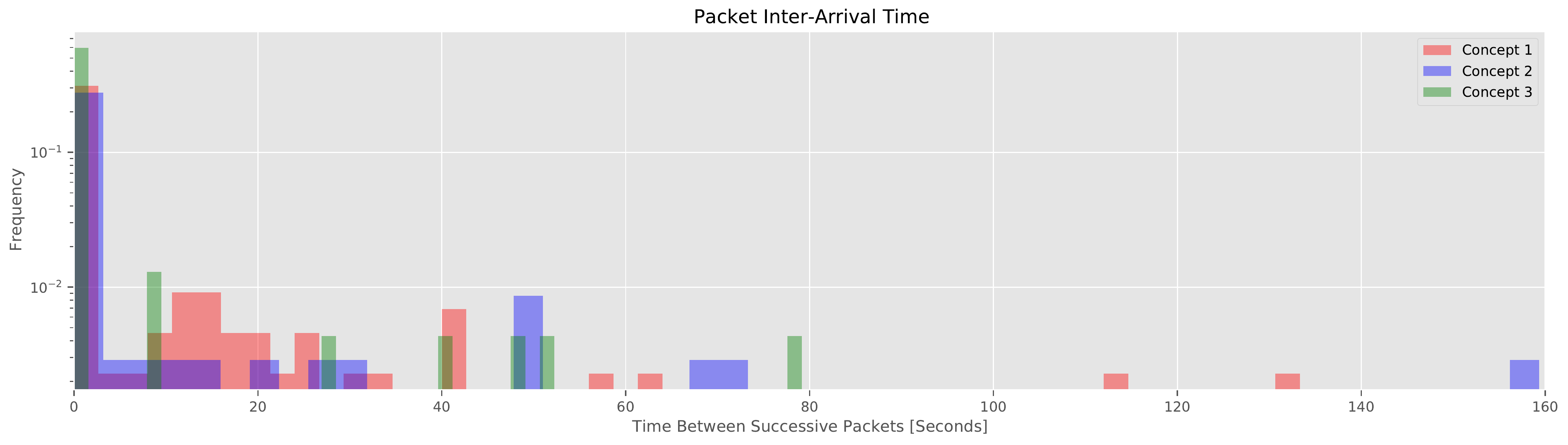}}
\caption{UE-to-Server Interaction Inter-Arrival Time}
\label{time}
\end{figure*}

While Fig. \ref{comparative} explores individual UE-to-server interactions from the three separate concepts, Fig. \ref{time} considers aggregate behaviour statistics of all UEs belonging to a single concept in terms of the packet inter-arrival time. This figure shows that the three concepts exhibit different characteristics in terms of packet arrival time, something which, as shown below, greatly impacts model performance. 

\subsection{Concept 2}
\begin{figure*}[!htbp]
\centerline{\includegraphics[width=1.95\columnwidth]{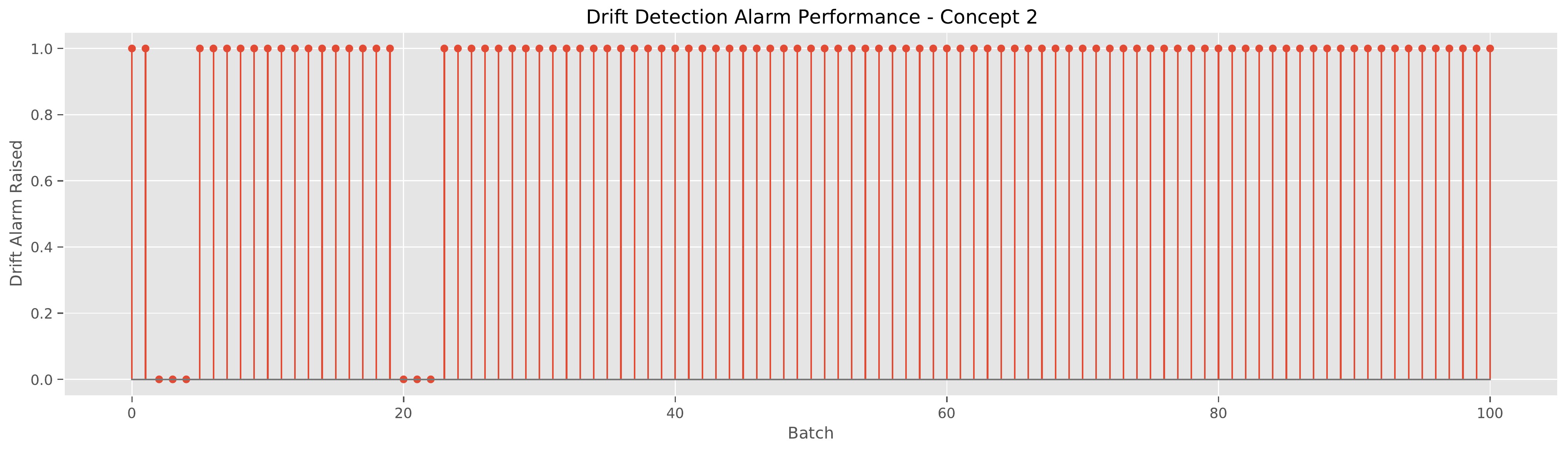}}
\caption{Drift Alarm Concept 2}
\label{DD_C2}
\end{figure*}

The following results will highlight the drift detection and adaptation framework’s performance on concept two. Figure \ref{DD_C2} presents the raised drift alarms for concept two without any drift adaptation. This serves as a baseline metric for performance as each time a value of one is assumed, a drift alarm has been raised. In theory, when using drift adaptation, the number of alarms should be reduced if the model is successfully able to adapt to the drift. This figure shows that only six of the considered batches did not raise a drift alarm. 

\begin{figure*}[!htbp]
\centerline{\includegraphics[width=1.95\columnwidth]{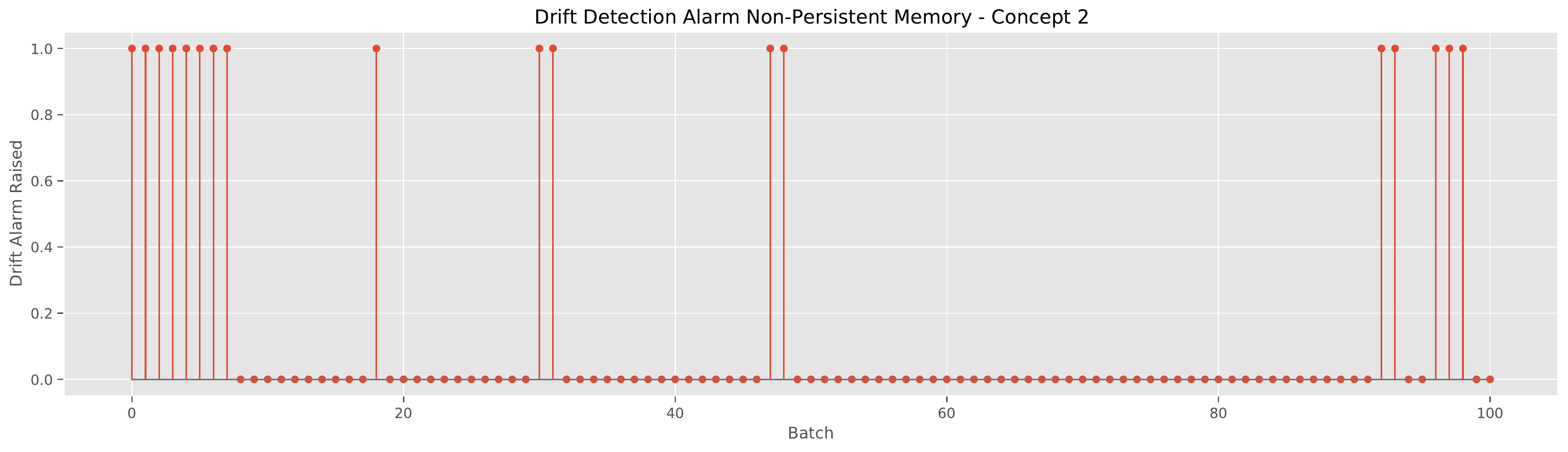}}
\caption{Non-Persistent Memory Mode Concept 2}
\label{DA_NP_C2}
\end{figure*}

To this end, Fig. \ref{DA_NP_C2} illustrates the raised drift alarms when using the non-persistent memory mode for the drift adaptation module. As seen in this figure, the number of raised drift alarms has been significantly reduced, demonstrating that the drift adaptation module is leading to the model successfully learning under the presence of drift. 

\begin{figure*}[!htbp]
\centerline{\includegraphics[width=1.95\columnwidth]{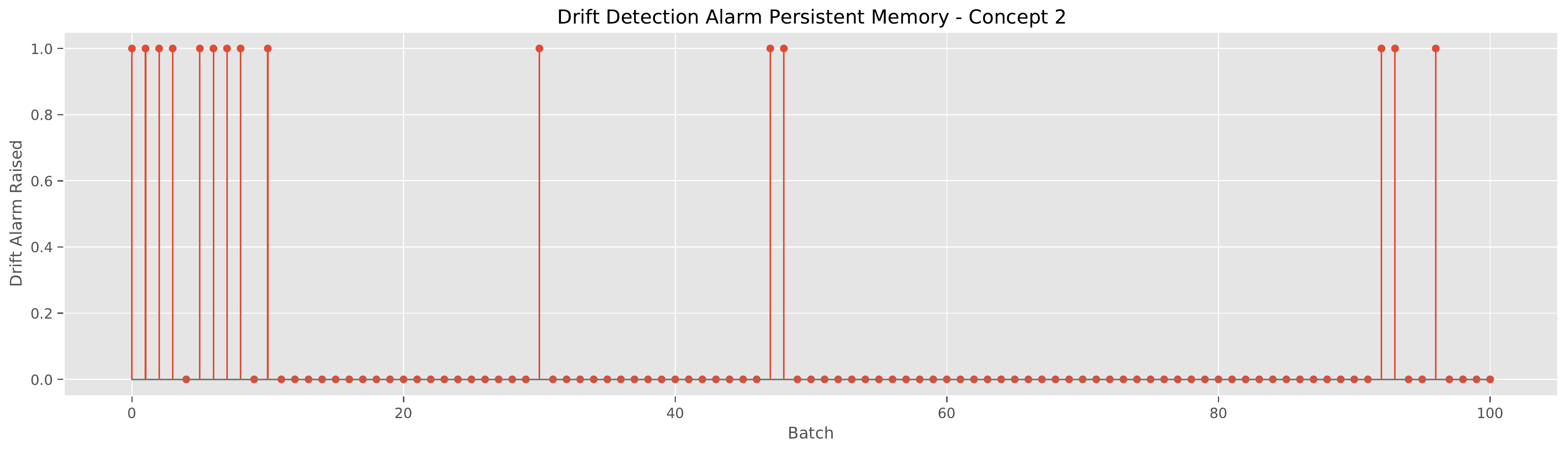}}
\caption{Persistent Memory Mode Concept 2}
\label{DA_P_C2}
\end{figure*}

Figure \ref{DA_P_C2} presents the drift alarms raised when using the persistent memory mode for the drift adaptation module. Comparing the performance of the persistent to the non-persistent memory mode, it is evident that the persistent memory mode results in fewer raised drift alarms. In this mode, it took the framework longer to adapt to the drift, evident by the higher number of drift alarms raised in the first few batches; however, once the model has adapted, the number of drift alarms raised is less compared to the non-persistent memory mode. 

As an additional metric of comparison, Eq. \ref{drift_red} presents the percentage of alarm reductions between the baseline comparison and each of the operational modes, where $n_b$ denotes the number of drift alarms raised in the baseline, and $n_a$ denotes the number of alarms raised in the compared mode.

\begin{equation}
\label{drift_red}
\frac{n_{b} – n_{a}}{n_{b}} \times 100 
\end{equation}

Furthermore, Eq. \ref{persist_red} is used to calculate the percentage of alarm reduction between the persistent and non-persistent memory modes, where $n_n$ denotes the number of alarms raised in the non-persistent memory method, and $n_p$ denotes the number of alarms raised in the persistent memory method.

\begin{equation}
\label{persist_red}
\frac{n_{n} – n_{p}}{n_{n}} \times 100 
\end{equation}

Considering these metrics, the non-persistent memory mode resulted in an 81\% alarm reduction compared to the baseline, and the persistent memory mode resulted in an 84\% alarm reduction. This translates to a 16.7\% improvement exhibited by the persistent memory mode compared to the non-persistent memory mode.

\subsection{Concept 3}
The following results will highlight the drift detection and adaptation framework’s performance on concept three. As previously discussed, the third concept is the most dissimilar to concept one and is expected to be much more difficult to adapt to. For the sake of brevity, the results of this drift will be discussed in terms of the previously defined comparison metrics. The non-persistent memory mode yielded a 32\% reduction in raised alarms compared to the 35\% reduction exhibited by the persistent memory mode. This translates to a 4\% improvement by the persistent over the non-persistent memory mode. As demonstrated by these results, despite being a much more difficult concept to learn, the drift detection and adaptation framework proposed results in a significant level of learning and a reduction of drift alarms.

\section{Conclusion}
The work presented in this paper considers a drift detection and adaptation framework for 5G core networks. Concept drift scenarios were emulated using a fully functional prototype of a 5G core network, and real network data was used as an evaluation method for the proposed framework. Results demonstrate a significant reduction of drift alarms raised by the system and a high degree of learning in the presence of drift. Future work in this area will consider the system’s parameter optimization as well as incorporate additional methods to mitigate severe drifts.

\bibliographystyle{unsrt}
\bibliography{sample}

\end{document}